# Shear-strain-induced two-dimensional slip avalanches in rhombohedral MoS$_2$

*Jing Liang[1,2], Dongyang Yang[1,2], Yunhuan Xiao[1,2], Sean Chen[1,2], Jerry I Dadap[1,2], Joerg Rottler[1,2] and Ziliang Ye[1,2]\**

[1]Department of Physics and Astronomy, The University of British Columbia, Vancouver, BC V6T 1Z1, Canada

[2]Quantum Matter Institute, The University of British Columbia, Vancouver, BC V6T 1Z4, Canada

**ABSTRACT**

Slip avalanches are ubiquitous phenomena occurring in 3D materials under shear strain and their study contributes immensely to our understanding of plastic deformation, fragmentation, and earthquakes. So far, little is known about the role of shear strain in 2D materials. Here we show some evidence of two-dimensional slip avalanches in exfoliated rhombohedral $MoS_2$, triggered by shear strain near the threshold level. Utilizing interfacial polarization in $3R-MoS_2$, we directly probe the stacking order in multilayer flakes and discover a wide variety of polarization domains with sizes following a power-law distribution. These findings suggest slip avalanches can occur during the exfoliation of 2D materials, and the stacking orders can be changed via shear strain. Our observation has far-reaching implications for developing new materials and technologies, where precise control over the atomic structure of these materials is essential for optimizing their properties as well as for our understanding of fundamental physical phenomena.

KEYWORDS: two-dimensional materials, slip avalanche, ferroelectricity, shear strain, mechanical exfoliation

Power law distributions are a common occurrence in nature and can provide significant insights into various phenomena[1-4]. In material systems, one well-known example of this is the critical behavior that occurs near a phase transition. Even when materials have different compositions and structures, they will share the same critical power-law exponent if they are part of the same universality class[5]. Beyond the phase transition at the thermal equilibrium limit, power law distributions and universality classes can also be found in non-equilibrium transitions[6], which often involve external driving forces. For example, when solids are stressed above a critical level, a cascade of strain drops can occur through isolated slip events, known as the slip avalanche phenomenon[7-21]. The size of these slips follows a power-law distribution in a wide range of three-dimensional solids ranging from nanocrystals to earthquakes, spanning over 12 orders of magnitude of length scale[19].

Two-dimensional (2D) materials have provided a unique platform for studying new science and technology at the atomic level. These materials, which are exfoliated from van der Waals (vdW) crystals, have a much weaker bonding force between atomic layers than the intralayer bond. As a result, out-of-plane compressive strain is widely used to tune the interlayer distance and coupling in these materials[22-24]. In contrast, in-plane shear strain can also be generated between layers in 2D materials, but its effect has not been previously investigated. In this study, we report the discovery of a power-law distribution of the polarization domain size in exfoliated molybdenum disulfide ($MoS_2$) with rhombohedral stacking order. We attribute this highly abnormal domain-size distribution among polarization domains to an interlayer slip avalanche induced by a shear strain near the threshold. Our findings suggest that slip avalanches can occur in 2D materials at the mechanical exfoliation stage, highlighting the potential for exploring the interlayer coupling of 2D materials in new ways.

2D materials with rhombohedral stacking order exhibit a range of fascinating emergent properties. For instance, in rhombohedral trilayer graphene, unconventional superconductivity has been observed in proximity to a metallic phase with reduced isospin symmetry[25,26]. Similarly, rhombohedral transition metal dichalcogenides (TMDs) exhibit asymmetric interlayer coupling, resulting in spontaneous out-of-plane polarization[27] that can be switched by an external electric field via in-plane sliding[28]. This unique property makes them attractive for a range of optoelectronic applications[29-32]. However, fully understanding the detailed stacking order in flakes of atomic thickness is a non-trivial task. As an example, in artificially stacked TMDs with marginal twists, the atomic lattice can naturally relax into a periodic array of nanoscale domains with alternating stacking orders[33-35]. In order to address such a challenge, we use electrical force microscopy (EFM) and Kelvin probe force microscopy (KPFM) to investigate the spontaneous polarization in rhombohedral $MoS_2$ flakes exfoliated from chemically synthesized 3R crystals[36,37].

EFM and KPFM are powerful techniques for characterizing polar materials with high spatial resolution[38]. The principle of applying them to probe the stacking configuration in 3R-$MoS_2$ is illustrated in Figure 1A, B. Taking bilayer 3R-$MoS_2$ as an example, two symmetric stacking configurations can coexist within a flake: AB stacking denotes the stacking configuration in which the sulfur atom in the top layer lies above the molybdenum atom in the bottom layer while BA stacking is the mirror image of the AB stacking configuration. Between the two stacking domains is a narrow domain wall region where one layer of $MoS_2$ is strained (Figure S1)[39]. In the AB stacking, since the spontaneous polarization has been determined to be upward[27], the resultant downward depolarization electric field leads to a higher surface potential. In the BA stacking, the polarization and depolarization field are reversed, giving rise to a lower surface potential (Figure 1B). Assuming the $SiO_2$ layer is sufficiently thin so the bottom layer in the entire flake has the same electric potential as the substrate[40], we find the total difference in the surface potential $V_{DC}$ between the two domains is twice the interlayer potential, $\Delta V_{DC} = 2Pd_0/\varepsilon_0\varepsilon_m$, where $P$ is the polarization strength, $d_0$ and $\varepsilon_m$ are the thickness and dielectric constant of monolayer MoS$_2$, $\varepsilon_0$ is the free-space permittivity. As the surface potential can induce an electrostatic force on an adjacent atomic force microscope (AFM) probe, the surface potential contrast between two types of domains can be visualized in the EFM mode. If a feedback voltage $V_{DC}$ is supplied to compensate the potential difference between the surface and tip, $\Delta V_{DC}$ can be quantitatively measured in the KPFM mode. In our experiment, the voltage is applied between the AFM tip and the highly conductive Si substrate (Figure 1A, Methods). Since the capacitance between the tip and sample is orders of magnitude smaller than that between the sample and substrate, the voltage drop across the SiO$_2$ is negligible (Supporting Text S1).

We first apply these two imaging techniques to a flake consisting of monolayer and bilayer regions (Figure 1C). After performing the EFM mapping, two types of domains with a clear electrostatic force contrast show up in the bilayer region, in stark contrast to the homogenous monolayer region. In the meantime, no corresponding contrast exists in the topography, indicating the signal originates purely from the stacking-induced surface potential difference (Figure S3). Compared to the artificial stack, the domains in the 3R-MoS$_2$ do not usually have obvious periodicity and often have much larger sizes. To quantify $\Delta V_{DC}$, we zoom in on a domain wall area and carry out the KPFM measurements (Figure 1D). The surface potential profile across the domain wall indicates a potential difference of $123 \pm 12$ meV, consistent with previous experimental reports[27,35]. We have also performed EFM and KPFM measurements in a 2H-MoS$_2$ flake with different thickness as a control (Figure S4), and they show little surface potential contrast. These findings confirm that the contrast observed in the EFM and KPFM images originate from the rhombohedral stacking configuration.

Compared to the bilayer case, the surface potential map of thicker 3R-MoS$_2$ flakes exhibits some interesting differences. First, the surface potential contrast across the domain wall is no longer single-valued. In a trilayer flake, four kinds of distinguishable surface potentials are observed in the EFM map (Figure 2A). (The contrast in the topography is a stitching artifact, as a single scan cannot cover the entire flake. A homogenous optical image of the flake is shown in Figure S5.) This result is interesting as a trilayer has two interfaces, which can bring four possible stacking configurations: CBA (↓, ↓), BAB (↓, ↑), ABA (↑, ↓), and ABC (↑, ↑), with arrows indicating the polarization direction at each interface. The CB and BC stacking are equivalent to the BA and AB stacking respectively. Among the four stacking configurations, BAB and ABA should give the same surface potential since the polarizations from adjacent interfaces are antiparallel to each other. However, the four observed surface potential levels suggest that the degeneracy between the ABA and BAB configurations is broken. Such a broken degeneracy in the surface potential among symmetric configurations is robust in experiments, as even more $\Delta V_{DC}$ values are observed in flakes thicker than the trilayer (an 8-layer result and analysis is shown in Figure S6,7).

After comparing our observations with previous KPFM studies in few-layer graphene[41,42], we conclude that the broken degeneracy can be explained by the substrate doping effect: As the SiO$_2$ substrate inevitably dopes electrons in the MoS$_2$ flake, free carriers tend to reside in the layer of lowest potential, which is different between BAB and ABA stacking configurations. As a result, the electric field between layers becomes modified, which ultimately changes the surface potential relative to the substrate. In Supporting Text S2, we have built a discrete Thomas-Fermi model to discuss such broken degeneracy in detail. According to this model, we color-code the EFM domains by their corresponding stacking configuration. Several domains with curved boundaries are discarded as they are unlikely intrinsic to the stacking configuration.

The second interesting aspect of the trilayer EFM map concerns the size of the polarization domain and its distribution. As shown in Figure 2B, two very large CBA yellow domains occupy over 90% of the flake and many small domains occupy the rest. In total, there are nearly 40 domains within this 40 x 40 $\mu m^2$ flake: Over thirty domains have less than 5 $\mu m^2$ in size while a few others are orders-of-magnitude larger (Figure 2C). Overall, these domains form a highly abnormal size distribution, which we do not expect to originate from the crystal growth, since a crystal grown from a single seed should not have such a high density of stacking fault[29,36] while a crystal grown out of multiple seeds should have a Gaussian distribution in domain size[43,44].

The distribution of the domain size becomes clear when the cumulative number of domains below a certain domain size $S$, $N(S)$, is plotted versus $S$ in a log-log plot. The cumulative plot is often used when the total sample number is not sufficient to directly perform statistics on the probability distribution[45]. In our experiments, tens of domains can be directly observed in a single homogenous flake. (Among multiple batches of samples, about 20% of the flakes exhibit such high domain densities.) As shown in Figure 2D, a linear fit can match all data points spanning four orders of magnitude in size, suggesting the cumulative domain number is approximately a power-law function of domain size: $N(S) \sim S^{-(\kappa-1)}$. In other words, the probability for a domain of certain size to appear is also power-dependent on the domain size with an exponent $\kappa$ of about 1.55 according to our fit (Table S1, Methods). Similar power-law distributions are observed in multiple flakes of different thickness (Figure S8). Although the detailed stacking configuration can no longer be individually assigned among thicker flakes, the parallel fit lines in the cumulative plot indicate different samples share a similar power-law exponent (Table S1).

As introduced in the beginning, the slip avalanches observed in a wide range of solid-state systems have a universal power-law exponent about 1.5[19], which is close to our fitted $\kappa$. We therefore interpret the experimental observation as an evidence of a two-dimensional slip avalanche during the mechanical exfoliation process triggered by an in-plane shear force (Figure 3A). Mechanical exfoliation has been an indispensable approach for preparing high-quality 2D materials. During exfoliation, an external force is imposed on the material surface mostly along the surface normal direction $F_\perp$ using some adhesive handle. If $F_\perp$ overcomes the vdW bonding force between layers, delamination happens, which stochastically leaves atomically thin flakes on the substrate. On the other hand, the exfoliation force unavoidably has in-plane components in practice $(F_\parallel)$[46,47]. Since the bottom of the remaining flake is attached to the substrate, $F_\parallel$ induces a shear strain, which can lead to slip avalanches, as observed in other three-dimensional systems[19].

Here we model the slip avalanche in 3R-MoS$_2$ as a cascade of domain wall movements that alter the stacking configuration. For simplicity, we only consider the tensile domain wall in a bilayer 3R-MoS$_2$ where domain wall lattice undergoes a tensile strain and the interlayer displacement direction is perpendicular to the domain wall (Figure 3B, S1A)[39]. Given the weak vdW bonding between two layers, an in-plane shear force can move the top layer leftwards, converting the AB domain near the domain wall to BA stacking[48]. Effectively, the domain wall moves rightward, antiparallel to the shear force direction. The domain wall of shear type, where the interlayer displacement direction is parallel to the domain wall, is expected to move perpendicular to the force direction with similar characteristics (Figure 3B, S1B).

A real 2D crystal is not perfect – an exfoliated flake can host a network of quenched disorders, which can hinder the domain wall movement, commonly known as pinning centers. These pinning centers can be atomic defects or localized residual strain. When the domain wall is pinned, a shear force overcoming the pinning potential can depin the domain wall and let it sweep through a portion of the flake until getting pinned again by stronger pinning center (Figure 3C). Collectively, a 2D network of randomly distributed pinning centers with varying potential strength can lead to a critical shear force threshold for the domain wall to sweep throughout the entire flake, according to the nonequilibrium depinning transition theory[49,50]. Right above the critical threshold, we expect some isolated domains are protected by strong pinning centers and remain unslipped, giving rise to a power-law distribution of domain sizes.

In light of this picture, we adopted a two-dimensional random field Ising model (RFIM) on a square lattice to simulate qualitatively the shear-force-induced domain wall movement[49-52]. Before the avalanche happens, the entire simulation domain has the AB stacking order, except for the leftmost column, as represented by the local polarization state $s = -1$ and $+1$ respectively. Every cell has a local slip threshold energy $E_i = J \sum_j s_i s_j + p_i$, where $s_j$ is the polarization of the nearest neighbor of the cell $s_i$, $J$ is the interaction energy between them, and $p_i$ is the local pinning potential. When a global external field is turned on, the $E_i$ of every cell is computed to check if the shear strain energy $\varepsilon$ is larger than the local threshold energy. The polarization of a cell is switched from $-1$ to $+1$ if the slipping criterion is met. The external field scanning is carried out column by column, left to right, to mimic the domain wall propagation effect. When the BA domain reaches the rightmost column, a domain wall percolation is achieved by the external field. More details about the RFIM simulation can be found in Supporting Text S3.

We first find the critical external field by simulating the RFIM model for a hundred times, each time with a randomly generated pinning potential network (Figure 3D). When the ratio of shear strain energy over interaction energy is over 1.2, the domain wall can sweep through the simulation space with a percolation probability close to 1. The critical field is found to be insensitive to the simulation space size or detailed pinning potential distribution (Figure S9A). Seven representative simulation results under different values of shear force are shown in Figure 3E and Figure S10. As predicted by the depinning transition theory, the domain wall either is stuck by pinning centers under weak shear forces (Figure S10A-C) or sweeps through the whole space under strong shear forces (Figure S10D-F). Only near the critical point (Figure 3E) do the domains in the simulation show similar shapes as in our experiments. More importantly, the statistics of the domain size at this condition also exhibits a power-law distribution with an exponent close to our experiments, as discussed below. The reason why multiple flakes exfoliated in different batches all experience similar slip avalanches near the critical point in our experiments might be biased by statistics – those that are away from the critical point exhibit little domain contrast and therefore cannot provide sufficient statistics on the domain size.

Next, we perform a statistical analysis on simulated domains in the vicinity of the critical point. Since all slipped units are connected as a single percolated domain, the isolated unslipped domains account for the majority of the statistics. As shown in Figure 3F, the domain size distribution follows a power law with an exponent of ~1.40, comparable to our experimental findings. The deviation from the power law distribution in the large and small domain limit is known to be caused by the size of the simulation space and the spatial resolution, respectively[19,53]. Similar to the critical field, the exponent is not very sensitive to simulation details (Figure S9B). In future studies, other factors such as a triangle lattice and associated anisotropy can be incorporated into a more comprehensive model. The fact that the ABC and CBA domains occurs more frequently than the ABA and BAB domain in our experiment may suggest certain stacking configurations are more energetically favorable than others.

The shear-force-induced interlayer sliding is further evidenced by the large-scale moiré superlattice occasionally observed in the exfoliated 3R-MoS$_2$. As the shear force is unavoidably inhomogeneous during exfoliation, the non-zero moment of the force can lead to a torque and therefore, an interlayer rotation (Figure 4A). Similar to twisted bilayer graphene, the rotational misalignment between layers gives rise to a moiré superlattice whose period $\lambda$ depends on the twist angle $\theta$, $\lambda = \frac{a}{2\sin(\theta/2)}$, where $a$ is the in-plane lattice constant of the monolayer MoS$_2$ (Figure 4A). If the twist angle is small, the energetically favorable AB- and BA-stacking domains are maximized through local lattice reconstruction, giving rise to a staggered triangle stacking domains[54,55]. Experimentally, we can image these reconstructed domains using KPFM (Figure 4B, S11). In our natural moiré superlattice, the observed period is about 200 nm, indicating the twisted angle is about 0.09°, much smaller than the intentionally created twists in artificial stacks. Such a large period and correspondingly small twist is also observed in another thick flake with rich domain contrast (Figure S12).

In this study, we have investigated the stacking order of exfoliated few-layer 3R-MoS$_2$ flakes by utilizing stacking-induced surface potential contrast. We have discovered a variety of polarization domains with a power-law distribution in size, which we attribute to shear-strain-induced two-dimensional slip avalanches at the mechanical exfoliation stage. Our findings suggest that shear strain is a promising way to tune the stacking order in TMDs, providing an opportunity to observe non-equilibrium statistical phenomena in two dimensions. Furthermore, it would be intriguing to explore whether slip avalanches can also occur in other 2D materials, such as graphene, since the percentage of rhombohedral domains in exfoliated few-layer graphene is similar to that in bulk graphite[56], and local strain has been shown to be capable of altering the stacking order of multilayer graphene[57]. Our study provides valuable insights into the fundamental understanding of the properties of 2D materials with rhombohedral stacking order and opens up new avenues for the design and application of these materials in various fields.

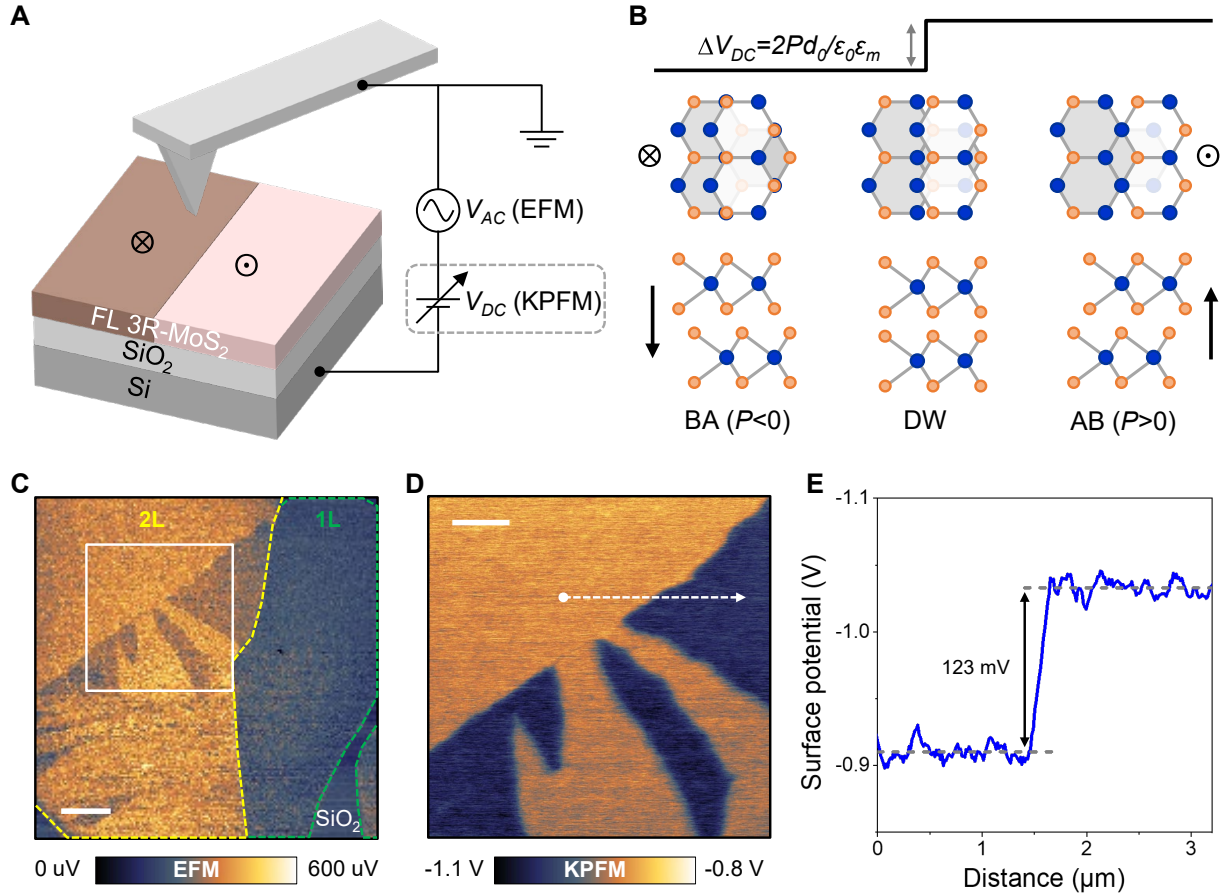

**Figure 1. Visualizing stacking configurations in 3R-MoS$_2$.** **(A)** Illustration of the experimental setup and the sample geometry. The Kelvin probe force microscopy (KPFM) measures the local surface potential $V_{DC}$ with a DC voltage as feedback. The electrostatic force microscopy (EFM) measures the electrostatic force between the tip and sample surface. **(B)** Top view (top row) and side view (bottom row) illustration of a 3R-MoS$_2$ bilayer with BA ($P < 0$), saddle-point (domain wall), and AB ($P > 0$) stacking configurations. The blue and orange circles denote the Mo and S atoms, respectively. For clarity, the bottom layer is shaded in grey. The corresponding polarization directions are marked by black arrows. **(C)** EFM mapping of a 3R-MoS$_2$ flake with both mono- and bilayer regions. Scale bar, 2 $\mu m$. **(D)** Surface potential mapping of the 3R-MoS$_2$ bilayer region enclosed by the white box in **(C)**. Scale bar, 1 $\mu m$. **(E)** Surface potential contrast measured along the dashed line in **(D)**, which is commensurate with the interlayer potential.

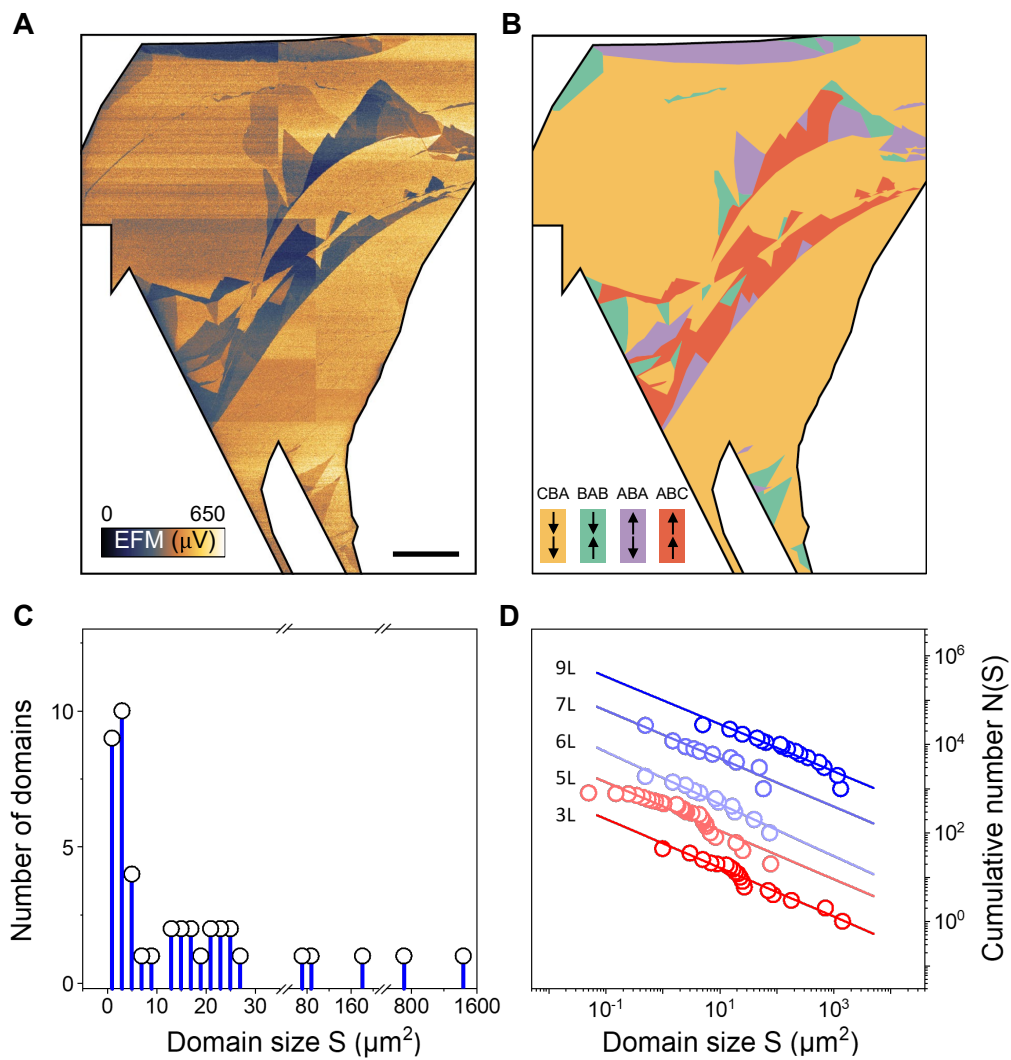

**Figure 2. Domain size distribution of a 3R-MoS$_2$ trilayer.** **(A)** EFM mapping of a 3R-MoS$_2$ trilayer. Scale bar, 10 $\mu m$. **(B)** Corresponding stacking configuration map of **(A)**. The yellow, green, purple, and red color represent four stacking configurations in trilayers as determined by the surface potential contrast: CBA, BAB, ABA, and ABC. **(C)** Histogram of the domain size, regardless of the specific stacking order. **(D)** Cumulative number $N(S)$ of domains with size larger than $S$ versus domain size S of 5 different samples with varying thicknesses. The solid line fit suggests the data follow a power-law dependence. To offset the data, the cumulative numbers of 5L, 6L, 7L and 9L samples are multiplied by 20, 100, 1000 and 1000 for clarity. The fitted power-law exponents $\kappa$ are summarized in Table S1.

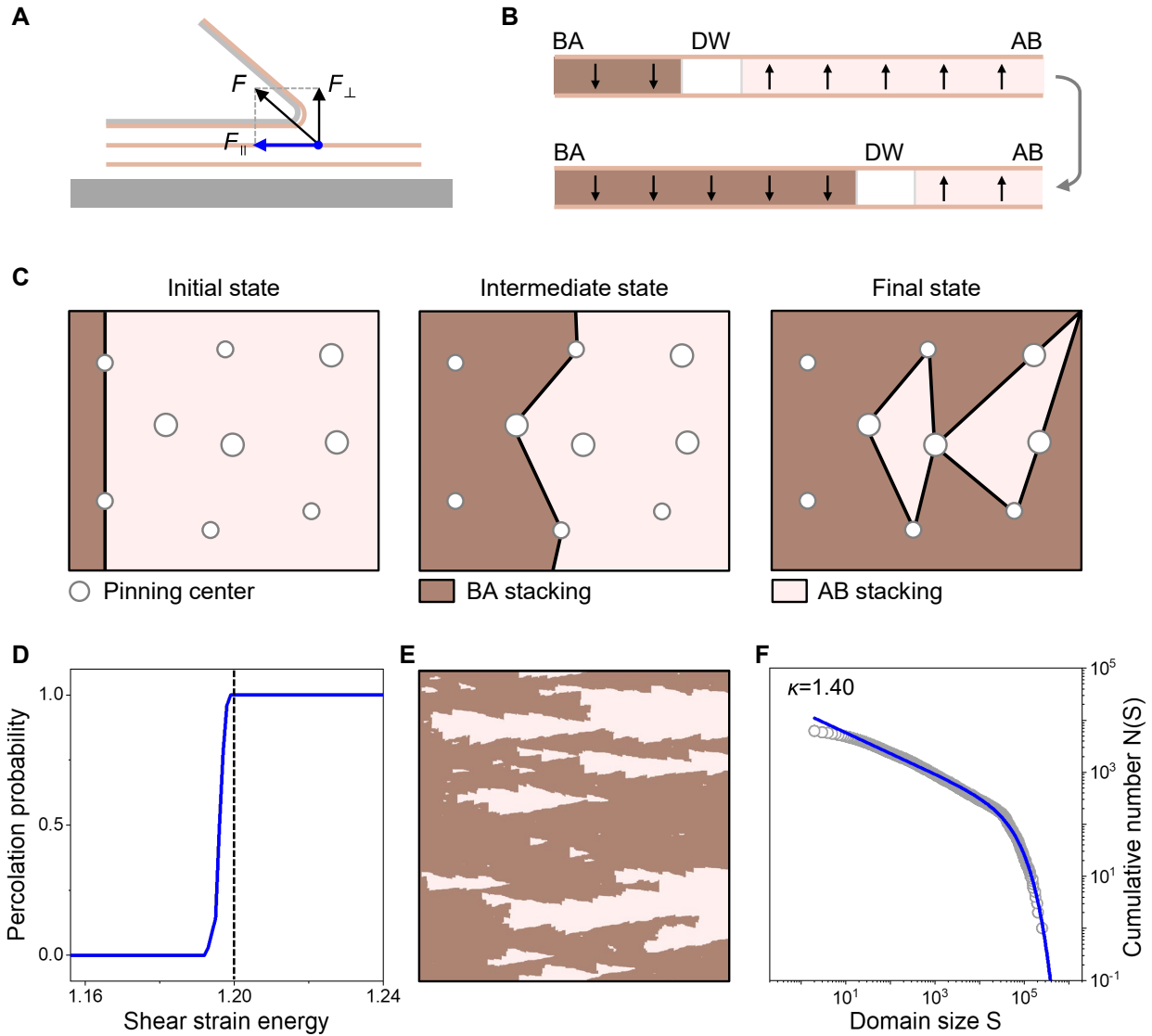

**Figure 3. Shear-force-induced two-dimensional slip avalanches.** **(A)** Schematic of the mechanical exfoliation process. Grey and pink parts are the exfoliation handle and MoS$_2$ layers. $F$ is the total force applied to the crystal surface. $F_\perp$ and $F_\parallel$ are the normal and shear force components, respectively. **(B)** Under a shear strain, the stacking order can be switched, which is equivalent to a domain wall (DW) moving rightward. The brown and pink regions represent the BA and AB domains while the arrow indicates the polarization direction. **(C)** Illustration of the slip avalanche process. When the shear strain energy is larger than the pinning potential, the domain wall becomes depinned until it is trapped by another pinning center with an even stronger pinning potential. A 2D network of strong pinning centers can keep some AB domains from being flipped. **(D)** The simulated probability for the BA domain to percolate the entire simulation space (900 × 900) under different shear strain energy. The dashed line denotes the critical shear strain energy for the depinning transition. **(E)** A representative simulation result corresponding to the shear strain energy at the dashed line condition, about 1.201. **(F)** The statistics of the simulated domain size at the same critical condition. The fit is based on the power-law dependence in addition to the exponential cutoff due to the limited simulation space.

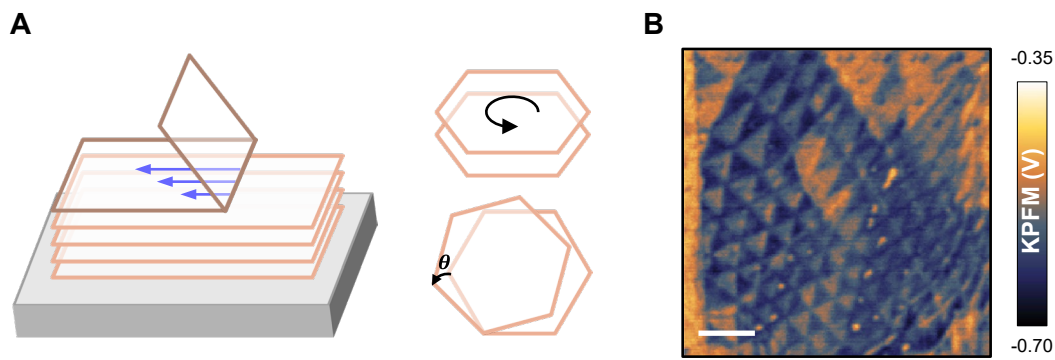

**Figure 4. Inhomogeneous shear-force-induced interlayer rotation. (A)** An inhomogeneous shear force applied to the 3R-MoS$_2$ surface can induce an interlayer twist. **(B)** KPFM mapping of a twisted 3R-MoS$_2$ flake induced by inhomogeneous shear force during the exfoliation. Scale bar, 400 $nm$.